\documentstyle[preprint,aps,a4,12pt]{revtex}
\tightenlines
\begin{document}
\draft
\hfill\vbox{\baselineskip14pt
            \hbox{\bf KEK-TH-619}
            \hbox{KEK Preprint 99-xxx}
            \hbox{February 1999}}
\baselineskip20pt
\vskip 0.2cm 
\begin{center}
{\Large\bf Theoretical modeling for quantum liquids from 1d to 2d 
dimensional crossover using quantum groups}
\end{center} 
\vskip 0.2cm 
\begin{center}
\large Sher~Alam\footnote{Permanent address: Department of Physics, University
of Peshawar, Peshawar, NWFP, Pakistan.}
,~M.~O.~Rahman\footnote{Photon Factory, KEK, Tsukuba, Ibaraki 305, Japan}
\end{center}
\begin{center}
{\it Theory Group, KEK, Tsukuba, Ibaraki 305, Japan }
\end{center}
\vskip 0.2cm 
\begin{center} 
\large Abstract
\end{center}
\begin{center}
\baselineskip=18pt
\noindent
 Recent experimental and theoretical work in strongly correlated
 electron system necessitates a formulation to deal with 1d to 2d 
 dimensional crossover and raises several interesting questions. 
 A particularly interesting question is what happens to the 1d Luttinger 
 liquid, as we go from 1d to 2d?
 The issue of dimensional crossover in lower dimensional systems
 is of great general interest, in particular the relationship 
 between quantum liquids and dimensional crossover from 1d to 2d.
 Thus one may ask given a 1d Luttinger liquid, what happens
 if we go from 1d to 2d, does the Luttinger liquid end up
 as a Fermi liquid or a non-Fermi liquid [NFL] in 2d. More importantly
 and relevantly what are the conditions that determine which
 type of quantum liquid we will end up with. Looking
 ahead to future applications we need to know what parameters
 to tune in order to engineer 2d quantum liquids from their 1d
 counterparts. In other words what are the parameters that
 determine quantum phase transitions. Can the Luttinger liquid
 lead to stripe formation in  high temperature superconducting
 [HTSC] materials in a natural way?
 Can we classify the quantum liquids existing in 1d, 2d and 3d
 under symmetry classes of quantum groups in practically
 meaningful way? What is the nature of dimensional phase
 transitions and how are these [i.e. dimensional phase transitions]
 made possible by interactions of quantum liquids. The main
 point of emphasis of the present note is that the transition
 from 1d to 2d has an underlying quantum group symmetry. The
 relationship of 1d to 2d transition with quantum group symmetry
 ties in nicely with our previous proposal to model superconductivity
 , antiferromagnetism and related phases arising from strongly
 correlated electron states with quantum groups. A simple
 model based on two interacting chains with nearest neighbor
 and next to nearest neighbor interactions is suggested.
 It appears reasonable to assume that it is the interaction
 between the 1d Luttinger liquids which leads to the 
 quantum group symmetry and the transition to 2d
 quantum liquid. The conditions under which this 2d
 quantum liquid deviates away from the 2d Fermi liquid
 is clearly of main interest to us. One of the prime motivations 
 underlying our proposal is the experimental observation 
 of {\em stripe} structure [phase] in high T$_{\rm c}$ superconductivity 
 [HTSC] materials. In short a simple model of two coupled chains
 1d chains is considered to answer an important question, how
 does one couple two 1d chains [Luttinger Liquid] to obtain NFL 
 or partial NFL behavior in 2d? 
\end{center}
\vfill
\baselineskip=20pt
\normalsize
\newpage
	
	The discovery of cuprate superconductors has initiated
an intense interest in strongly correlated system. In particular
the anomalous normal state behavior of these materials indicate
strong deviations from normal Fermi Liquid behavior. A very
useful and classic behavior non-Fermi liquid [NFL] is provided
by one-dimensional [1d] electron gas, where the generic fixed
point behavior is believed to be a Luttinger liquid \cite{hal81}.
Exact solutions of 1d systems are available via Bethe ansatz
and its generalizations. There is a wide array of non-perturbative
techniques one can use in 1d. The `unique' behavior of the 1d Luttinger 
liquid fixed points can be traced back to or originates in special 
kinematics of 1d. The Fermi surface in 1d consists of just two
points, the electrons interact very strongly at these two points.
Moreover the energy and momentum conservation impose a single
constraint on scattering processes, resulting in an enhancement
in scattering phase space. Simply, the `limited' phase space 
available in 1d for scattering is qualitatively responsible for the 
peculiar behavior of Luttinger liquid. It is naturally interesting
to see if the peculiar behavior of 1d Luttinger liquid may
carry over to 2d. Anderson \cite{and90} made an hypothesis
about the form of the ground state of the 2d Hubbard model
by identifying singular scattering diagrams. Thereafter
several authors have tried to generalize the Luttinger
liquid concepts to higher dimensions \cite{cas91,hou93,net95,met98}.

	It is clear that the special kinematics of 1d will not 
be present in 2d and other higher dimensions. A straightforward 
generalization of Luttinger liquid to higher dimensions cannot be 
expected since the 1d kinematics impose a single constraint
on scattering processes whereas in higher dimensions the 
energy and momentum [kinematics] impose several independent
constraints on the scattering process. Thus one finds that
these constraints lead to the stabilization of the Fermi
liquid in 2d and higher dimensions \cite{cas91,sha94}.
An interesting approach to go from 1d to 2d which suggest
itself would be to couple 1d Hubbard chains since the Hubbard 
Hamiltonian [HH] and its extensions 
dominate the study of strongly correlated electrons systems and the 
insulator metal transition. Indeed Lin et al.~\cite{lin97}
have taken such an approach and have claimed that in
2d one ends up with Fermi liquid. So how does one
avoid or bypass the Fermi liquid as one goes from 1d to 2d.
One way is by the introduction of long-range or singular
interactions \cite{and90,bar93,kim94}. Yet another approach
is based on Anderson's \cite{and90-1} ingenious suggestion that
NFL behavior in higher dimensions may arise from the formation
of bound anti-bound states above and below the single particle
continuum. This idea has been further examined very recently
by Ho and Coleman \cite{ho99}. Despite these efforts a
route to NFL behavior in 2d that uses strictly local
interactions has not been reported.  

	In an attempt to gain a fundamental understanding
of the relationship between antiferromagnetism and superconductivity
and other quantum phases [as mentioned above] that may be present in 
strongly correlated electron systems we are naturally led to examine
the issue of the nature of the dimensional crossover from 1d to 2d
and 2d to 3d. In the present note we concentrate on the
1d to 2d case. The main point of emphasis of the present note is that 
the transition from 1d to 2d has a underlying quantum symmetry group. 
This relation of 1d to 2d transition with quantum group symmetry ties 
in nicely with our previous proposal to model superconductivity, 
antiferromagnetism and related phases arising from strongly correlated 
electron states with quantum groups \cite{ala98,ala99}.
We thus adopt the following general strategy:
\begin{itemize}
\item{}We first identify a 1d model such as 1d Hubbard Hamiltonian.
At this stage we allow ourself the freedom to choose among the
plethora of known 1d Hamiltonians and/or guess any reasonable 1d 
Hamiltonian which could form the building block of our final 2d model 
for HTSC and antiferromagnetism. We initially restrict our 1d Hamiltonian
to be integrable for simplicity however our real goal is to relax
the condition of integrability to suit a particular realistic 
situation of interest.
\item{}We next couple two 1d systems and try to identify
the relevant 2d `covering' model.
\end{itemize}

	We write the Hamiltonian for the two coupled chains
interacting via the nearest neighbor and next nearest neighbor
interactions as
\begin{eqnarray}
H &=& \sum_{i}t[\sigma_{i}^{x}\sigma_{i+1}^{x}+
\sigma_{i}^{y}\sigma_{i+1}^{y}+\xi \sigma_{i}^{z}\sigma_{i+1}^{z}]\nonumber\\
&&+\sum_{j}t[\tau_{j}^{x}\tau_{j+1}^{x}
+\tau_{j}^{y}\tau_{j+1}^{y}+\xi \tau_{j}^{z}\tau_{j+1}^{z}]\nonumber\\
&&+\sum_{i,j}t^{'}[\sigma_{i}^{x}\tau_{j}^{x}
+\sigma_{i}^{y}\tau_{j}^{y}+\xi \sigma_{i}^{z}\tau_{j}^{z}]\nonumber\\
&&+\sum_{i,j}t^{''}[\sigma_{i}^{x}\tau_{j+1}^{x}
+\sigma_{i}^{y}\tau_{j+1}^{y}+\xi \sigma_{i}^{z}\tau_{j+1}^{z}]\nonumber\\
&&+\sum_{i}U\sigma_{i}^{z}\tau_{i}^{z},
\label{h1}
\end{eqnarray}
where $t$ is the nearest neighbor hopping parameter
on the same chain, $t^{'}$ is the nearest neighbor
hopping parameter on different chains and
$t^{''}$ is the next to nearest neighbor hopping
parameter on different chains. In Eq.~\ref{h1} we
have assumed su(2) symmetry and written the
Hamiltonian in bosonized form. $\sigma$ and $\tau$ 
are two independent sets of Pauli matrices. 
The Hamiltonian in Eq.~\ref{h1} resembles that of two 
XXZ chains interacting with each other via nearest and next
to nearest neighbor interactions. The model Hamiltonian
in Eq.~\ref{h1} is written in bosonized form. 
It is well-known in 1d bosons and fermions are closely
related and that this relationship breaks down in
higher dimensions. 
If we set $t^{'}=t^{''}=\xi=0$ in Eq.~\ref{h1}, we obtain 
the bosonized form of the su(2) Hubbard Model,
\begin{eqnarray}
H = \sum_{i}t[\sigma_{i}^{x}\sigma_{i+1}^{x}+
\sigma_{i}^{y}\sigma_{i+1}^{y}]
+\sum_{j}t[\tau_{j}^{x}\tau_{j+1}^{x}
+\tau_{j}^{y}\tau_{j+1}^{y}]
+\sum_{i}U\sigma_{i}^{z}\tau_{i}^{z}.
\label{h2}
\end{eqnarray}
The fermionic version of Eq.~\ref{h2} is well-known
and can be immediately written down\footnote{We note
that $\sigma$ when it appears as a subscript is a spin label 
in Eq.~\ref{h3} and other fermionized Hamiltonians. It is not 
to be confused $\sigma$ used for Pauli matrices. Similar
remarks apply to $\tau$.}
\begin{eqnarray}
H = \sum_{i,\sigma}t[c_{i,\sigma}^{\dagger}c_{i+1,\sigma}+
c_{i+1,\sigma}^{\dagger}c_{i,\sigma}]
+\sum_{i}U[n_{i\uparrow}-\frac{1}{2}]
[n_{i\downarrow}-\frac{1}{2}].
\label{h3}
\end{eqnarray}
In Eq.~\ref{h3} we have used the following common
notation: The canonical Fermi operators satisfy the 
anticommutation relations: $\{c_{i,\sigma}^{\dagger}
c_{j,\sigma^{'}}\}=\delta_{ij}\delta_{\sigma\sigma^{'}}$
where $i,j=1,2,....,N$ and $\sigma,\sigma^{'}=\uparrow,\downarrow$. 
At any lattice site $i$ there are four electronic
states, namely: $|0\rangle$, $|\uparrow\rangle_{i}
=c_{i,\uparrow}^{\dagger}|0\rangle$, $|\downarrow\rangle_{i}
=c_{i,\downarrow}^{\dagger}|0\rangle$, and
$|\uparrow\downarrow\rangle_{i}
=c_{i,\uparrow}^{\dagger}c_{i,\downarrow}^{\dagger}|0\rangle$.
As usual we take $n_{i,\sigma}=c_{i,\sigma}^{\dagger}c_{i,\sigma}$ 
to denote the number operator for electrons at site $i$ and 
spin $\sigma$. It is straightforward to see that after summing 
over $\sigma$ at site $i$ one has 
$n_{i}=n_{i,\uparrow}+n_{i,\downarrow}$.

The fermionic version of Eq.~\ref{h1} can be written
as
\begin{eqnarray}
H &=& \sum_{i,\sigma}t[c_{i,\sigma}^{\dagger}c_{i+1,\sigma}+
c_{i+1,\sigma}^{\dagger}c_{i,\sigma}
+\xi (n_{i,\sigma}-\frac{1}{2})(n_{i+1,\sigma}-\frac{1}{2})]\nonumber\\
&&+\sum_{j,\tau}t[d_{j,\tau}^{\dagger}d_{j+1,\tau}+
d_{j+1,\tau}^{\dagger}d_{j,\tau}
+\xi (n_{j,\tau}^{'}-\frac{1}{2})(n_{j+1,\tau}^{'}-\frac{1}{2})]\nonumber\\
&&+\sum_{i,\sigma,j,\tau}t^{'}[c_{i,\sigma}^{\dagger}d_{j,\tau}+
d_{j,\tau}^{\dagger}c_{i,\sigma}
+\xi (n_{i,\sigma}-\frac{1}{2})(n_{j,\sigma}^{'}-\frac{1}{2})]\nonumber\\
&&+\sum_{i,\sigma,j,\tau}t^{''}[c_{i,\sigma}^{\dagger}d_{j+1,\tau}+
d_{j+1,\tau}^{\dagger}c_{i,\sigma}
+\xi (n_{i,\sigma}-\frac{1}{2})(n_{j+1,\sigma}^{'}-\frac{1}{2})]\nonumber\\
&&+\sum_{i,\sigma,j,\tau}(U/2)\{[n_{i,\sigma}-\frac{1}{2}]
[n_{i\tau}^{'}-\frac{1}{2}]+[n_{j,\sigma}-\frac{1}{2}]
[n_{j\tau}^{'}-\frac{1}{2}]\}.
\label{h4}
\end{eqnarray}
$d_{j,\tau}$ and $d_{j,\tau}^{\dagger}$ are 
canonical Fermi operators, with
$n_{j,\sigma}^{'}=d_{j,\tau}^{\dagger}d_{j,\tau}$.
 
We next want to construct a Hamiltonian which is a 
partially q-deformed version of the Hamiltonian in 
Eq.~\ref{h4}. To this end we define a deformation
function $D_{q}$ [see, for example, Ref.~\cite{bra97}
and references therein]
\begin{eqnarray}
 D_{q}&=& \exp\{-\frac{1}{2}(\eta-\sigma\gamma)n_{i,-\sigma}
 -\frac{1}{2}(\eta+\sigma\gamma)n_{i+1,-\sigma}\}\nonumber\\
q &=& \exp\{\gamma\}\nonumber\\
 \exp\{-\eta\} &=& \frac{q^{\alpha+1}-q^{-\alpha-1}}
{q^{\alpha}-q^{-\alpha}}
\label{h5}
\end{eqnarray}
where $\alpha \in {\rm C}$ is a free parameter.
We note the usual definition used in quantum groups
\cite{kli97}
\begin{eqnarray}
[x]_{q} \stackrel{\mathrm{def}}{=}
\frac{q^{x}-q^{-x}}{q-q^{-1}},~~~~~~~x \in {\rm C}.
\label{h6}
\end{eqnarray}
The next step is to q-deform the Hamiltonian given
in Eq.~\ref{h4}. We note that many possibilities
exist if we want to partially q-deform, since
we can q-deform either chain or parts which are
responsible for the interaction between the chains.
For simplicity we just consider the deformation
of one of the chain, viz
  \begin{eqnarray}
H_{q} &=& \sum_{i,\sigma}t[c_{i,\sigma}^{\dagger}c_{i+1,\sigma}+
c_{i+1,\sigma}^{\dagger}c_{i,\sigma}
+\xi (n_{i,\sigma}-\frac{1}{2})(n_{i+1,\sigma}-\frac{1}{2})]
D_{q}\nonumber\\
&&+\sum_{j,\tau}t[d_{j,\tau}^{\dagger}d_{j+1,\tau}+
d_{j+1,\tau}^{\dagger}d_{j,\tau}
+\xi (n_{j,\tau}^{'}-\frac{1}{2})(n_{j+1,\tau}^{'}-\frac{1}{2})]\nonumber\\
&&+\sum_{i,\sigma,j,\tau}t^{'}[c_{i,\sigma}^{\dagger}d_{j,\tau}+
d_{j,\tau}^{\dagger}c_{i,\sigma}
+\xi (n_{i,\sigma}-\frac{1}{2})(n_{j,\sigma}^{'}-\frac{1}{2})]\nonumber\\
&&+\sum_{i,\sigma,j,\tau}t^{''}[c_{i,\sigma}^{\dagger}d_{j+1,\tau}+
d_{j+1,\tau}^{\dagger}c_{i,\sigma}
+\xi (n_{i,\sigma}-\frac{1}{2})(n_{j+1,\sigma}^{'}-\frac{1}{2})]\nonumber\\
&&+\sum_{i,\sigma,j,\tau}(U/2)\{[n_{i,\sigma}-\frac{1}{2}]
[n_{i\tau}^{'}-\frac{1}{2}]+[n_{j,\sigma}-\frac{1}{2}]
[n_{j\tau}^{'}-\frac{1}{2}]\}.
\label{h7}
\end{eqnarray}
There are several directions in which we can generalize
the basic Hamiltonian model considered in the
present note, namely
\begin{itemize}
\item{}From both physical and purely technical point
of views we may consider its supersymmetric variants
in particular with pair hoppings \cite{ese92,ese93}.
\item{}The q-deformed versions of the supersymmetric
variants \cite{bra97,arn97}.
\item{}We may readily generalize the current model
to include the effect of doping. In particular
we may begin with the spin-1 version of Hamiltonian
considered in this note doped with spin-1/2
carriers \cite{fra98}.
\end{itemize}

	Next we detail some background. It is well-known that 
the Hubbard Hamiltonian [HH] has an su(2) symmetry \cite{fra91}, this 
is especially apparent if we write the HH in terms of two independent 
sets of Pauli matrices $\sigma$ and $\tau$ 
\begin{eqnarray}
H^{HH}=\sum_{i}t[\sigma_{i}^{x}\sigma_{i+1}^{x}+
\sigma_{i}^{y}\sigma_{i+1}^{y}+\tau_{i}^{x}\tau_{i+1}^{x}+
\tau_{i}^{y}\tau_{i+1}^{y}]+U\sigma_{i}^{z}\tau_{i}^{z}.
\label{h1x}
\end{eqnarray}
It is convenient to consider writing the HH in terms
of some basic `building blocks'. It is straightforward
and well-known that a building block of HH is the XX model
with Hamiltonian
\begin{eqnarray}
H^{XX}=\sum_{i}t[\sigma_{i}^{x}\sigma_{i+1}^{x}+
\sigma_{i}^{y}\sigma_{i+1}^{y}].
\label{h2x}
\end{eqnarray}
In turn the XX model can be regarded as a particular
case of the XXZ model 
\begin{eqnarray}
H^{XXZ}=\sum_{i}t[\sigma_{i}^{x}\sigma_{i+1}^{x}+
\sigma_{i}^{y}\sigma_{i+1}^{y}+\xi \sigma_{i}^{z}\sigma_{i+1}^{z}],
\label{h3x}
\end{eqnarray}
when $\xi=0$. So far we have assumed su(2) symmetry. It
is well-known that exact integrability is a feature of several
toy models of statistical mechanics, namely 2d Ising model,
six-vertex model, eight-vertex model, and others. 
A 2d classical statistical model which is a `covering' model for
1d Hubbard model has been identified by Shastry \cite{sha86}
\footnote{We note that in the present paper we have considered
su(2) Lie groups as one of the basic ingredient.
The generalization to other Lie groups is an
interesting issue and has been dealt by several
authors, for example the su(N) XX model and its
variant has been considered by Maassarani and
Mathieu \cite{maa97}. In context of integrable
supersymmetric and q-deformed supersymmetric
models there are several works, for example see 
Ref.~\cite{arn97} and references therein}.  
In particular it was shown in ~\cite{sha86} that the 2d
covering model of the 1d HH provides a one parameter
family transfer matrices which commute with the HH and
that two transfer matrices of a family mutually commute.
Underlying the commutation relation is the Yang-Baxter
factorization condition. In turn underlying the Yang-Baxter 
relations and knots are quantum groups \cite{kak91,kli97}.
We must be careful to distinguish between the classical
Yang-Baxter relations from their quantum counterparts.
The classical Lie group underlies classical Yang-Baxter
equation just as the quantum group underlies the
quantum Yang-Baxter equation. Indeed this is one
way of defining quantum group from its classical
Lie group counterpart. We caution the 
reader that currently there is no `satisfactory' general 
definition of a quantum group\footnote{We mean in terms of rigorous
mathematics}. However it is commonly accepted \cite{kli97}
that quantum groups are certain `well-behaved' Hopf
algebras and that the standard deformations of the enveloping
Hopf algebras of semisimple Lie algebras and of coordinate Hopf
algebras of the corresponding Lie groups are guiding examples.
An amazing feature of quantum group theory is the 
unexpected connections with apparently unrelated
concepts in physics and mathematics such as Lie Groups,
Lie Algebras and their representations, special functions,
knot theory, low-dimensional topology, operator algebras,
noncommutative geometry, combinatorics, quantum inverse
scattering problem, theory of integrable models, conformal
and quantum field theory and perhaps others. In summary
a class of non-commutative Hopf algebra was found in the 
investigations of integrable systems. In turn these
Hopf algebras are q-deformed function algebras of
classical groups, this structure can be taken to define
a quantum group. The structure of quantum groups further
suggest that one may envision the possibility of even
discarding the commutativity of the algebra of coordinate
functions. The new class of symmetry based on the 
noncommutativity of the algebra of coordinate functions 
may have applications to real physical systems 
[such as HTSC and related phases in strongly
correlated electron systems] other than the {\em integrable} 
systems. Indeed we want to go beyond integrable systems. 
However it is first useful to understand the exact
role of quantum groups in modeling of HTSC and
other quantum phases and what happens to interacting
integrable 1d models as they evolve towards 2d. 
One of the prime motivations underlying our proposal
is the experimental observation of {\em stripe} structure [phase]
in HTSC materials. A number of experimental techniques have 
recently observed that the ${\rm CuO_{2}}$ are rather inhomogeneous, 
providing evidence for 
phase separation into a two component system. i.e. carrier-rich 
and carrier-poor regions. In particular, extended x-ray absorption 
fine structure [EXAFS] demonstrated that these domains forms stripes 
of undistorted and distorted local structures alternating with 
mesoscopic length scale comparable with coherence length in HTSC.
In theories based on magnetic interactions \cite{and87}
for modeling of HTSC, it has been assumed that the CuO$_{2}$
planes in HTSC materials are microscopically homogeneous.
However, a number of experimental techniques have recently observed 
that the CuO$_{2}$ are rather inhomogeneous, providing evidence for 
phase separation into a two component system. i.e. carrier-rich and 
carrier-poor regions \cite{oyn99}. In particular, extended x-ray 
absorption fine structure [EXAFS] demonstrated that these domains 
forms stripes of undistorted and distorted local structures alternating 
with mesoscopic length scale comparable with coherence length in HTSC.
The neutron pair distribution function of Egami et al. \cite{ega94}
also provides structural evidence for two component charge carriers.
Other techniques also seem to point that below a certain temperature
T$^{*}$ \footnote{The following can be taken as a definition of T$^{*}$:
T$^{*}$ is an onset temperature of pseudogap opening in spin or
charge excitation spectra.} the CuO$_{2}$ planes may have ordered stripes
of carrier-rich and carrier-poor domains \cite{ega94}. The emergence
of experimental evidence for inhomogeneous structure has led to
renewal of interest, in theories of HTSC which are based on 
alternative mechanism, such as phonon scattering, the lattice
effect on high T$_{\rm c}$  
superconductivity \cite{lat92,phs93,phs94,ega94}.
Polarized EXAFS study of optimally doped YBa$_{2}$Cu$_{3}$O$_{\rm y}$
shows in-plane lattice anomaly \cite{oyn99} below a characteristic 
temperature T$^{*'}$\footnote{T$^{*'}$ may be defined as follows: 
T$^{*'}$ is an onset temperature of local phonon anomalies
and T$^{*'} < T^{*}$.} which lies above T$_{\rm c}$, and
close to the characteristic temperature of spin gap opening
T$^{*}$. It is an interesting question if the in-plane
lattice anomaly is related to the charge stripe or spin-phonon
interaction. We note that it has been attempted in \cite{fuk92,fuk93}
to relate the spin gap observed in 
various experiments such as NMR, neutron scattering and transport
properties to the short-range ordering of spin singlets.   

	In the context of engineering of quantum liquid behavior
it is interesting to note the work by Bianconi et al., \cite{bia98}.
In Ref.~\cite{bia98} a model of stripes is constructed in the
context of HTSC materials. In this model \cite{bia98} a
gas of free electrons with effective mass $m^{*}$ moves
in a superlattice of quantum stripes of width $L$
separated by a periodic potential barrier. Here \cite{bia98}
the 2d striped phase is formed by superconducting 
stripes of width $L$ [U-stripes] alternated by
separating stripes of width $W$ [D-stripes]. Their
conclusion \cite{bia98} which is of interest to 
us is that the maximum critical temperature is reached at the 
cross-over from 2d to 1d behavior. Indeed this observation
is in agreement with our intuitive remark in \cite{ala98},
namely ``superconductivity arises when two immiscible
phases namely 2-D antiferromagnetic state and a 3-D
metallic state, are ``forced'' to meet at 
$T_{c}\rightarrow \infty$.'' 

	In conclusion we have proposed a model which
can lead to a theory of dimensional crossover from
1d to 2d. We have also presented a partially q-deformed 
version of the model Hamiltonian. It is expected that
this model can also be used to study the underlying physics
of spin ladder compounds \cite{bal98}.
\section*{Acknowledgments}
The S.~Alam's work is supported by the Japan Society for
the Promotion of Science [JSPS]. S.~Alam thanks Prof.~H.~Oyanagi
for introducing him to stripes and Profs.~K.~Hagiwara,~H.~Kawai
and T.~Yukawa, A.~Bianconi,~H.~Oyanagi and others for useful
discussions.  
\\
\\
\\
\\
\\

Note Added:-We think it is our duty scientific or
otherwise to make the following remarks regarding
citations and references. It is a fact nowadays
that most authors do not bother to cite other
works until that work belongs to a particular
group or groups that they are affiliated with in 
one way or the other. A lot of times we tried to 
cite work which could be relevant to our work. But given 
the present climate of citations we think that is unfair 
to us. Moreover one simply cannot go through many works
since there are so many. We have thus decided that we are 
not obligated just like most to consume our time going 
through other peoples work. In summary we can disadvantage 
ourself unnecessarily by going through other people work
who do not care to cite work of others beside
themselves and their favorite groupings. However in 
the present case we have followed up our usual policy 
of following up on references as much as possible. 


\end{document}